\documentclass[prb,a4paper,showpacs,superscriptaddress,twocolumn]{revtex4}
\usepackage{amssymb}
\usepackage{amsmath}
\usepackage{amsfonts}
\usepackage{graphicx}
\usepackage{bm}
\usepackage{color}
\usepackage{float}
\usepackage{verbatim}
\usepackage{epstopdf}
\usepackage{caption}

\renewcommand{\Im}{\,\textrm{Im}\,}
 \DeclareMathOperator{\Tr}{Tr}

\definecolor{DarkBlue}{rgb}{0,0,0.6}

\def\ve{\varepsilon}

\def\bn{\mathbf{n}}

\def\bA{\mathbf{A}}

\def\br{\mathbf{r}}

\def\bp{\mathbf{p}}
\def\bn{\mathbf{n}}

 1

\begin{document}

\title{Effect of disorder on the transverse magnetoresistance of Weyl semimetals}

\author{Ya.~I.~Rodionov}
\affiliation{Institute for Theoretical
and Applied Electrodynamics, Russian Academy of Sciences, Izhorskaya str. 13, Moscow, 125412 Russia}
\affiliation{National University of Science and Technology MISIS, Moscow, 119049 Russia}
\affiliation{National Research University Higher School of Economics, Moscow, 101000 Russia}

\author{K.~I.~Kugel}
\affiliation{Institute for Theoretical and Applied Electrodynamics, Russian Academy of Sciences, Izhorskaya str. 13, Moscow, 125412 Russia}
\affiliation{National Research University Higher School of Economics, Moscow, 101000 Russia}

\author{B.~A.~Aronzon}
\affiliation{P.~N.~Lebedev Physical Institute, Russian Academy of Sciences, Moscow, 119991 Russia}

\author{Franco~Nori}
\affiliation{Theoretical Quantum Physics Laboratory, RIKEN Cluster for Pioneering Research, Wako-shi, Saitama 351-0198, Japan}
\affiliation{Physics Department, The University of Michigan, Ann Arbor, MI 48109-1040, USA}

\begin{abstract}
We study the effect of random potentials created by different types of impurities on the transverse magnetoresistance of Weyl semimetals. We show that the magnetic field and temperature dependence of the magnetoresistance is strongly affected by the type of impurity potential. We analyze in detail two limiting cases: ($i$) the ultra-quantum limit, when the applied magnetic field is so high that only the zeroth and first Landau levels contribute to the magnetotransport, and ($ii$) the semiclassical situation, for which a large number of Landau levels comes into play. A formal diagrammatic approach allowed us to obtain expressions for the components of the electrical conductivity tensor in both limits. In contrast to the oversimplified case of the $\delta$-correlated disorder, the long-range impurity potential (including that of Coulomb impurities) introduces an additional length scale, which changes the geometry and physics of the problem. We show that the magnetoresistance can deviate from the linear behavior as a function of magnetic field for a certain class of impurity potentials.
\end{abstract}

\pacs{72.10.-d
%Theory of electronic transport; scattering mechanisms
72.80.-r
%Conductivity of specific materials
75.47.-m
%Magnetotransport phenomena; materials for magnetotransport
5.60.Gg
%Quantum transport
}

\date{\today}

\maketitle

\section{Introduction}
\label{intro}

In recent years, problems related to quantum transport in materials with the Dirac spectrum of charge carries, in particular in Weyl semimetals, have attracted  considerable interest~\cite{HosurCRPh2013,LuShenFrPh2017}. Much effort was focused on the longitudinal magnetoresistance, where the negative contribution associated with the so-called chiral anomaly arising due the transfer of charge carries between Weyl points with opposite chiralities plays a dominant role~\cite{SpivakPRB2013,BurkovPRL2014,HuangPRX2015,ShekhtarNatPhys2015,LiNatCom2015,SpivakPRL2018}.

At low magnetic fields, rather nontrivial manifestations of the weak localization and antilocalization effects have also been addressed~\cite{LuPRB2015,LuChinPh2016}. No less interesting is the behavior of the transverse magnetoresistance, where a nonsaturating linear magnetic field dependence is observed at high fields~\cite{LiangNatMat2015,FengPRB2015,NovakPRB2015,ZhaoPRX2015}. The nature of such unusual behavior has been widely discussed.

The main physical mechanisms in the ultra-quantum regime were revealed in the seminal work of Abrikosov~\cite{AbrikosovPRB1998}. He considered a gapless semiconductor with a linear dispersion law near the chemical potential. The chemical potential itself coincides with the zeroth Landau level and the charge carriers are scattered by impurities characterized by a screened Coulomb potential.

This problem was generalized in the detailed studies presented in Refs.~\onlinecite{KlierPRB2015,KlierPRB2017,PesinPRB2015}, which were  stimulated by numerous experimental observations of the linear magnetoresistance. In these papers, the main emphasis was on the case of point-like impurities.

In the opposite limit of long-range impurity potentials, treated by the Born and self-consistent Born approximations, it was possible to obtain only qualitative results. The results for an arbitrary position of the chemical potential were obtained in Ref.~\onlinecite{PLeePRB2017} within the framework of a similar approximation; and also in Ref.~\onlinecite{PLeePRB2015}, where the approach based on the classical motion of the so-called guiding center was used. For the screened Coulomb potential of impurities, the electron transport was also analyzed in the case of a gapped Dirac spectrum~\cite{KonyePRB2018,KonyePRB2019}.

Thus, we see that in spite of these serious efforts, some important aspects of the analysis of the magnetoresistance of Weyl semimetals remain untouched. First of all, its behavior at different characteristic ranges of the impurity potential, and accurate calculations in the limit of long-range potentials. These issues still remain unsolved even in the most well-studied ~\cite{AbrikosovPRB1998,KlierPRB2015,KlierPRB2017} ultra-quantum limit, when only one Landau level contributes to the magnetoresistance.

In this paper, we first focus on the calculations in the ultra-quantum limit and calculate components of the magnetoconductivity tensor based on the accurate diagrammatic approach which we have formulated earlier~\cite{RodionovPRB2015}. Then, we present a detailed analysis of the semiclassical limit, when a large number of Landau levels contributes to the transport characteristics.

In Section~\ref{Model}, we formulate the model and introduce all the necessary parameters. In Section~\ref{Ultraquantum}, we analyze the components of the electrical conductivity tensor and their magnetic field dependence in the ultra-quantum limit, for which the dominant contribution comes from the zeroth Landau level. In Section~\ref{Semiclassical}, we consider the magnetotransport at the semiclassical limit, for which the temperature is high enough and a large number of Landau levels comes into play. Both in Sections~\ref{Ultraquantum} and \ref{Semiclassical}, we put the main emphasis on the magnetotrasport for rather long-range impurity potentials. In Section~\ref{Discussion}, we discuss the obtained results. The details of our calculations are presented in the Appendix.

\section{Model and characteristic parameters}
\label{Model}

Our study is aimed at the analysis of the transport characteristics of the Weyl semimetal (WSM) with impurities under the effect of an applied transverse magnetic field (i.e., the magnetic field direction is perpendicular to that of the electric current). We start from the low-energy Hamiltonian for the WSM in its conventional form
\begin{gather}
    \label{ham1}
    \begin{split}
       H&=H_0+H_{\rm imp},\\
       H_0&= v\int\psi^\dag(\br)\bm{\sigma}\left(\bp-\frac{e}{c}\bA\right)\psi(\br) d\br,\\
       H_{\rm imp} &= \int \psi^\dag(\br) u(\br)\psi(\br) d\br,
    \end{split}
\end{gather}
where $H_0$ is the Hamiltonian of non-interacting Weyl fermions and  $H_{\rm imp}$ describes the interactions with the impurity potential;
$\bm{\sigma}=(\sigma_x,\ \sigma_y,\ \sigma_z)$ are the Pauli matrices acting in the pseudospin space of Weyl fermions, $\bp=-i\nabla_\br$ is the  momentum operator, $v$ is the Fermi velocity, and $u(\br)$ is the impurity potential.

The impurity potential is understood to be of a general form. It is of electrostatic origin, but its specific form as well as the form of its correlation function can be arbitrary. Of particular importance to the experiment is the screened Coulomb impurity potential. As was argued in Ref.~\onlinecite{SyzranovPRB2015}, there exists a regime, in which the Coulomb impurity scattering dominates over the electron--electron interaction (see the corresponding discussion in Section~\ref{Discussion}).

Throughout the paper, we set $\hbar = k_B = 1$. We also neglect the influence of different Weyl cones on each other, concentrating on the low-energy physics. The vector potential of the magnetic field \textbf{H} is chosen in the asymmetric gauge
\begin{gather}
    \label{vector-potential}
   \mathbf{A}=(0,Hx,0).
\end{gather}

\begin{figure}[h]
\centering
 \includegraphics[width=0.33\textwidth]{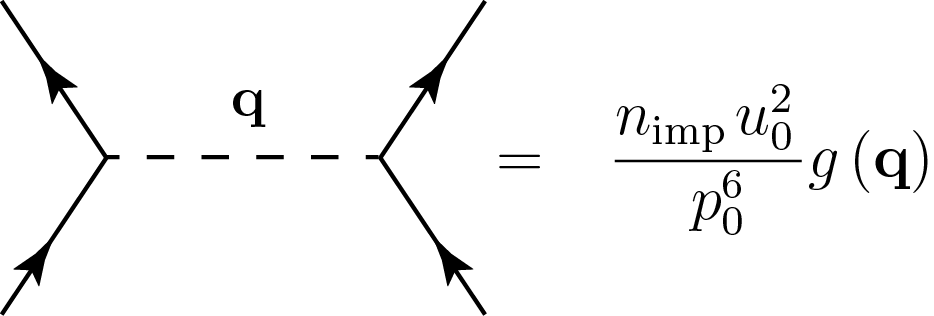}
 \caption{Disorder vertex for the perturbation theory. Here, the dashed line represents the disorder correlation function, while solid tails are fermion lines.}
\label{disorder0}
\end{figure}

In this paper, we will use the Kubo-type diagrammatic approach. The impurity potential thus enters the formalism in terms of its correlation function. The relevant diagram is shown in Fig.~\ref{disorder0}. We write the corresponding correlation function in the following form:
\begin{gather}
\begin{split}
  \int d\br e^{-i\bp\br}\langle u(\br)u(0)\rangle &\equiv g(\bp) =\frac{n_{\rm imp}u_0^2}{p_0^6}g_0\left(\frac{p^2}{p_0^2}\right),\\
  g_0\left(\frac{p^2}{p_0^2}\right)& =|u^2_\bp|\frac{p_0^6}{u_0^2}.
\end{split}
\end{gather}
Here, $n_{\rm imp}$ is the concentration of impurities and $u_0$ is the characteristic amplitude of the impurity potential. The disorder correlation function is written in terms of the dimensionless function $g$, which is introduced in momentum space from the very beginning. Also, $u_\bp$ is the Fourier transform of the impurity potential, while the  disorder
correlation length is $r_0=p_0^{-1}$. The $p_0^6$ factor is introduced from dimensional considerations.

In this work, we are focusing on the long-range correlation between disordered impurities (long-range disorder). In the ultra-quantum case
\begin{gather}
\max\{T,\mu\}\ll v/l_H,
\end{gather}
where
\begin{gather}
l_H=\sqrt{c/(eH)}
\end{gather}
is the magnetic length  and $\mu$ is the doping level of the WSM sample, and this limit corresponds to the condition:
\begin{gather}
\label{scales1}
   l_H \ll r_0 \ll \lambda,
\end{gather}
where
\begin{gather}
\lambda =v/\{\max{T,\mu}\}
\end{gather}
is the characteristic particle wavelength. Let us also introduce the energy scale associated with the magnetic field strength,
\begin{gather}
\Omega=v\sqrt{2eH/c},
\end{gather}
characterizing the distance between the zeroth and first Landau levels (LLs).
In the opposite semiclassical limit $\Omega \ll T$, the corresponding condition for the disorder correlation length reads:
\begin{gather}
\label{scales2}
   \lambda \ll r_0 \ll l_H.
\end{gather}
Limit~\eqref{scales1} intuitively appeals to the physical picture where the center of the magnetic orbit moves along the impurity potential line, while limit~\eqref{scales2} corresponds to the proper particle motion along the impurity potential line.

\section{Magnetotransport at $\{T,\mu\}\ll \Omega$ (ultra-quantum limit)}
\label{Ultraquantum}

\subsection{Computation of $\sigma_{xx}$}
\label{sigma_xx}

As was mentioned above, the ultra-quantum limit corresponds to the condition \begin{gather}
\Omega\gg \max\{T,\mu\}.
\end{gather}
This means that the first Landau level is high enough and only the ground state (and the first excited state) contributes to the magnetotransport. The $xx$ component of the conductivity tensor is determined by the following analytical expression:
\begin{gather}
\begin{split}
\label{cond-zero}
   &\sigma_{xx}=
      e^2v^2\int \frac{d\ve\; d\bp \; dx^\prime}{(2\pi)^3}\frac{d f(\ve)}{d\ve}\\
   &\times{\rm Tr} \langle {\rm Im}G^R_{11}(x,x^\prime;\ve,\bp){\rm Im}G^R_{22}(x^\prime,x;\ve,\bp) \rangle,
    \end{split}
\end{gather}
where angular brackets denote the averaging over disorder, $f$ is the Fermi distribution function, and the retarded Green's functions are defined as follows:
\begin{gather}
\label{green-zero}
    \begin{split}
    &G^R(x,x^\prime;\ve,\bp)= \sum\limits_{n=0}^\infty S_n(x_{p_y})G(\ve,\bp) S^\dag_n(x_{p_y}^\prime)\\
              &S_n(s)=\begin{pmatrix}
            \chi_n\big(s\big)\ &\ 0\\
            0\ &\ \chi_{n-1}\big(s\big)
           \end{pmatrix},\\
           &G(\ve,\bp)=\frac{\ve+v\bm{\sigma}\cdot \mathbf{p}_n}{(\ve+i0)^2-\ve_n^2}, \\
           &x_{p_y}=x-p_y l_H^2.\ \
    \end{split}
\end{gather}
Here, $\chi_n\big(s\big)$ is the oscillator normalized wave function of the $n$th state,
\begin{gather}
\mathbf{p}_n=(0,\sqrt{2n}/l_H,p_z)
\end{gather}
is the effective momentum, $\mathbf{p}_{yz}$ is the two-dimensional momentum $(p_y, p_z)$. We are using perturbation theory, therefore assuming that the concentration of impurities is not too high. The dimensionless expansion parameter characterising the disorder strength is assumed to be small:
\begin{gather}
    \label{expansion}
    \frac{1}{\ve\tau}\sim\frac{n_{\rm imp}u_0^2}{v^2p_0^5}\ll 1,
\end{gather}
where $\ve$ is the characteristic energy scale for charge carriers and $\tau$ is its impurity scattering time (see below Eq.~\eqref{times0}).

\begin{figure}[h]
\centering
 \includegraphics[width=0.4\textwidth]{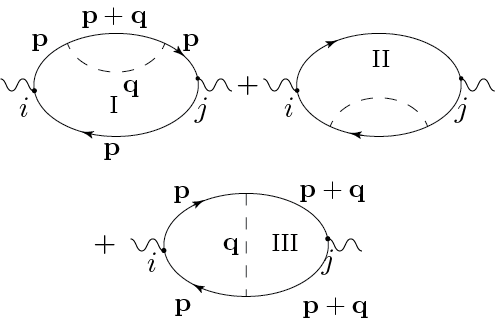}
 \caption{Three contributions to the conductivity in first-order perturbation theory. The disorder vertices are represented by dashed lines defined in Fig.~\ref{disorder0}. The indices $i,j$ become $x$ or $y$ depending on the type of conductivity which is computed.}
\label{conduct0}
\end{figure}

As the analysis shows, in the ultra-quantum limit, it is enough to keep the first order of perturbation in the disorder strength. Therefore, only three possible diagrams contribute to the conductivity (see  Fig.~\ref{conduct0}). Even less trivial is the fact that the vertex correction (diagram III in Fig.~\ref{conduct0}) is exponentially suppressed. Therefore, only the first-order disorder corrections to the Green's function are needed to be taken into account. This was first stated explicitly for the case of short-range disorder in~\cite{KlierPRB2015}.

In the ultra-quantum limit and in the long-range disorder case $p_0l_H \ll1$ (Eq.~\eqref{scales1}), only the zeroth and first LLs in Fig.~\ref{conduct0} contribute to the conductivity. For the Green's function $G^R_{11}$ in~\eqref{cond-zero}, it is enough to take only the contribution of the zeroth LL,
\begin{gather}
{\rm Im}\,G_{11}(\ve,\bp)=-\pi\delta(\ve-p_zv).
\end{gather}
The expression for the conductivity $\sigma_{xx}$ then goes along the same lines for the general type of disorder as that in Ref.~\onlinecite{AbrikosovPRB1998}.
The result is given by the following integral:
\begin{gather}
\label{conduct-uq}
  \sigma_{xx} =
\frac{e^2v^2}{4\Omega^2}
        \int\frac{dq_xdq_y}{(2\pi)^2}(q_x^2+q_y^2)
       g_{2,q_{xy}},
\end{gather}
where $g_{2,q_{xy}}$ is the effective two-dimensional disorder correlation function defined as
\begin{gather}
    \label{potential2D}
  g_{2,p_{xy}}=\int g(\bp)\frac{dp_z}{2\pi}\equiv g(p_{xy})\Big|_{z=0}.
\end{gather}
In Ref.~\onlinecite{AbrikosovPRB1998}, Eq.~\eqref{conduct-uq} was analyzed only in the case of the Coulomb impurity potential. We, however, come to the conclusion that for different types of disorder, the formula abovegives a qualitatively different $H$ dependence.

Before we proceed, let us make the following observation. The integral in Eq.~\eqref{conduct-uq} of the 2D disorder correlation function determining the conductivity can become divergent at high momenta (short-range case).
However, in  our calculations, we used the long-range disorder approximation, implying that $ql_H \ll 1$, where $q$ is the characteristic disorder momentum. Therefore, $q\sim l_H^{-1}$ is the natural short-range cutoff scale. As we will see below, the system exhibits different types of magnetotransport depending on the short-range behavior of the disorder, $p_0\ll q\ll l^{-1}_H$.

\subsection{$\sigma_{xx}$ for different short-range behavior of the impurity potential}

Let us now assume that the impurity potential has the short-range asymptotics
\begin{gather}
    u(r)=\frac{u_0}{(p_0r)^{1+\gamma}},\ \ l_H\ll r\ll p_0^{-1},\ \ -1<\gamma<1.
\end{gather}
Here, the natural constraint $\gamma<1$  means that we are not considering pathological cases of potentials leading to the ``falling to the center" phenomenon. On the other hand, the $\gamma < -1$ constraint should  exclude the unphysical case of decaying at $r=0$. Then, the disorder correlation function in momentum space reads:
\begin{gather}
    \label{potent}
    g(p)=n_{\rm imp}
        \begin{cases}
            l_H^4v^2,\ \ &p\ll p_0,\\
            \\
            u^2_0/p_0^6\left(p_0/p\right)^{4-2\gamma},\ \
              &p_0 \ll p \ll l^{-1}_H.
        \end{cases}
\end{gather}

The question we now address is: what is the behavior of the conductivity as a function of $H$ for different values of $\gamma$?
To answer this question, we analyze expression~\eqref{conduct-uq} for various cases discussed below.

($a$) $-1<\gamma<0.$
We call this the ``regular disorder" case. The integral in~\eqref{conduct-uq} is convergent and the convergence region is $p\sim p_0$. In this case, we have
\begin{gather}
    \label{sigmaxx_regular}
    \sigma_{xx} = \frac{ec}{16\pi H p_0^2}n_{\rm imp} u_0^2g_1,\quad \gamma<0,
\end{gather}
where $g_1=\int_0^\infty g(x^2)x^3dx$, with $x=p/p_0$, is a numerical constant, which depends on the details of the shape of the disorder distribution function.

As we are going to see below, the behavior corresponding to Eq.~\eqref{sigmaxx_regular} is identical to that characteristic of the Coulomb disorder, $\gamma=0$.

($b$) $\gamma=0$. For the Coulomb disorder, the integral determining the conductivity in ~\eqref{conduct-uq} is log-divergent. In the Coulomb case, the inverse Debye radius reads:
\begin{gather}
p_0=\sqrt{\alpha}l_H^{-1}
\end{gather}
for the case $\{T,\mu\}\ll \Omega$, where
\begin{gather}
\alpha=e^2/(\hbar v\kappa)
\end{gather}
is the WSM fine structure constant and  $\kappa$ is the permittivity. One recovers the result~\cite{AbrikosovPRB1998}:
 \begin{gather}
    \sigma_{xx} = \frac{e^3c\pi }{H} n_{\rm imp}\ln\frac{1}{\alpha},\quad \gamma=0.
\end{gather}

($c$)  $0<\gamma <1$.
We call this the ``singular disorder" casedue to its short-range behavior. The integral in~\eqref{conduct-uq} is then divergent at high momenta $q$ and an appropriate cut-off $p\sim l_H^{-1}$ should be introduced. In this case, we have a nontrivial result for $\sigma_{xx}$:
\begin{gather}
    \sigma_{xx} = \frac{ec }{16\pi H p_0^2} n_{\rm imp}u_0^2\left(\frac{eH}{cp_0^2}\right)^{\gamma},\quad 0< \gamma < 1.
\end{gather}
The above results can be summarized by the following formula:
\begin{gather}
    \label{sigmaxx}
  \sigma_{xx}=\begin{cases} e^2v^2p_0^4 g_1(c/eH) \sim H^{-1}, \quad -1<\gamma<0,\\
  \\
  e^2v^2\alpha^2\ln(1/\alpha)(c/eH) \sim H^{-1}, \quad \gamma=0, \\
  \\
  e^2v^2n_{\rm imp}u_0^2\left(\frac{eH}{cp^2_0}\right)^{\gamma}(c/eH) \sim H^{\gamma-1},\quad 0 < \gamma < 1.
  \end{cases}
\end{gather}
We see that the $H$ dependence of the conductivity is affected by the nature of the disorder. In particular, if the correlation function has stronger than Coulomb power-law growth at short distances, the corresponding exponent $\gamma$ enters the conductivity.

The parameter most relevant to many experiments is the magnetoresistance. To calculate it, we need to know the Hall conductivity $\sigma_{xy}$.

\subsection{Hall conductivity $\sigma_{xy}$}
The Hall conductivity is given by the sum of two terms:
\begin{gather}
    \label{sigmaxy_full}
  \sigma_{xy}=\sigma^{\rm I}_{xy}+\sigma^{\rm II}_{xy}.
\end{gather}
The first term in~\eqref{sigmaxy_full}, $\sigma^{\rm I}_{xy}$, is the so-called normal contribution, which is given by the following relation \cite{Streda}:
\begin{gather}
    \label{sigmaxy_norm}
    \begin{split}
      &\sigma^{\rm I}_{xy}=\frac{e^2\Omega^2}{4\pi^2}
      \int\frac{d\ve}{2\pi}\frac{df(\ve)}{d\ve}\\
      &\times\sum\limits_n
      \left[G^R_{22}\Im G^R_{11}\!-\!G^R_{11}\Im G^R_{22}\!-\! \Im G^R_{22} G^A_{11}\!+\! \Im G^R_{11} G^A_{22}\right].
    \end{split}
\end{gather}
As is seen from Eq.~\eqref{sigmaxy_norm}, it comes from the vicinity of the Fermi surface, as it is proportional to $df/d\ve$. In the absence of disorder, it is easily verified that $\sigma^{I}_{xy}=0$. Therefore, it is perturbative in the disorder strength. The second term in Eq.~\eqref{sigmaxy_full} is the so-called anomalous contribution.
It is proportional to the derivative of the charge carrier density with respect to the applied magnetic field $H$ and, as such, comes from the entire volume inside the Fermi surface. As is understood from the definition of the anomalous part
\begin{gather}
    \label{sigma-anom}
    \begin{split}
    \sigma_{xy}^{\rm II}&=ec\frac{dN(H,\mu,T)}{dH},\\
    N(H,\mu)&=\int_{-\infty}^{\infty}\nu(\ve)f_{\ve-\mu}d\ve,
    \end{split}
\end{gather}
it is nonzero even in the absence of disorder. Here, the density of states reads:
\begin{gather}
    \label{DOS1}
   \begin{split}
    \nu(\ve)={\rm Tr}\int\frac{dp_{yz}}{(2\pi)^2}{\rm Im}\,G(\ve,p_y,p_z,x,x)\\
    =\frac{1}{2\pi l_H^2}\sum\limits_{n}\int\frac{dp_z}{2\pi}{\rm Im}G_n(\ve,p_z).
    \end{split}
\end{gather}
Thus, from perturbative arguments, we understand that $\sigma_{xy}=\sigma^{\rm II}_{xy}$, i.e. it is determined by the anomalous part.

In our case (long-range disorder), it is even possible to compute $\sigma^{\rm II}_{xy}$ at all orders of perturbation theory in the strength of the disorder, in the limit $p_0^{-1}\rightarrow\infty$.
The result is independent of the disorder strength and is given by the disorder-free expression:
\begin{gather}
    \label{sigmaxy}
    \sigma_{xy}=\sigma^{\rm II}_{xy}=\frac{e^2\mu}{4\pi^2 v}=e^2n_0l_H^2,
\end{gather}
where $n_0$ is the charge-carrier density.
In experiments, $n_0$ is usually a fixed parameter stemming from the charge-neutrality condition due to the imbalance of donor and acceptor impurities in WSMs. Therefore, to compare with experiments, one needs to express the chemical potential in terms of $n_0$.

The formula for the resistivity is as follows:
\begin{gather}
    \rho_{xx}=\frac{\sigma_{xx}}{\sigma_{xx}^2+\sigma_{xy}^2}.
\end{gather}

Taking into account theexpressions for $\sigma_{xx}$~\eqref{sigmaxx} and $\sigma_{xy}$ ~\eqref{sigmaxy}, we obtain the following results for the field dependence of the resistivity in the ultra-quantum limit:

\begin{gather}
    \label{resistxx}
    \rho_{xx}\sim\begin{cases}H,\quad \quad -1<\gamma<0,\\
        H,\quad \quad \gamma = 0\quad{\rm Coulomb\ disorder},\\
      H^{1+\gamma},\ \ 0<\gamma<1,
    \end{cases}
\end{gather}
at fixed $n_0$. The expression~\eqref{resistxx} is an important result of our paper. It shows that \textit{measuring $\rho_{xx}(H)$ of the WSM in the ultra-quantum regime, one can extract information about disorder correlations and the form of the impurity potential}.

\section{Magnetotransport at $T\gg \Omega$ (semiclassical limit)}
\label{Semiclassical}

The opposite limit, which allows for an analytical treatment, is when the temperature of the WSM is much larger than $\Omega$. Here, we focus on the most experimentally viable case when the magnetic length isbeing much larger than the disorder correlation length
\begin{gather}
  \label{cond-p0}
    l_H^{-1}\ll p_0 \ll \frac{\max\{T,\mu\}}{v}.
\end{gather}
In the case of theCoulomb potential, $p_0$ is the inverse Debye screening length, and the right-hand side condition in Eq.~\eqref{cond-p0} is equivalent to $\alpha\ll1$ (which is true for a typical WSM, like $\rm{Ca}_2\rm{As}_3$, see Refs.~\onlinecite{Liu2014,Jay-Gerin1977}, where the $\alpha$ value is estimated~\cite{RodionovPRB2015} as $\sim0.05$). The left-hand side condition in~\eqref{cond-p0} in this case should be substituted by
\begin{gather}
  \label{cond-C}
    \Omega \ll \sqrt{\alpha}\max\{T,\mu\}.
\end{gather}
Therefore, the temperatures should not be too low.

\subsection{Computation of $\sigma_{xx}$}

Here, we need to keep in mind that the characteristic number of LLs contributing is large
\begin{gather}
      n\sim \max\{T,\mu\}/\Omega\gg1.
\end{gather}
This allows us to use the large $n$ asymptotics for LL wave functions $\psi_n(s)$. Their highly oscillatory behavior allows one to drastically simplify the calculations.

In the Appendix, we argue that in the $T\gg\Omega$ limit, the problem of disorder averaging becomes essentially a two-dimensional one. Unlike the ultra-quantum case, the corresponding 2D plane is $y=0$  (due to asymmetry of our vector potential gauge~\eqref{vector-potential}) with the effective correlation potential:
\begin{gather}
    \label{potential2Da}
  g_{2,p_{xz}}=\int g(\bp)\frac{dp_y}{2\pi p_0}\equiv g(p_{xz})\Big|_{y=0}.
\end{gather}

The most crucial observation is that all the integrals entering the Green's functions and Dyson equations are essentially orthogonality equations sometimes \textit{spoiled} by the potential enveloping function.
All the details of the perturbative analysis are summarized in the Appendix.
Despite the breakdown of translational invariance (the Green's functions depend on $x$ and $x^\prime$ separately rather than on $x-x^\prime$), it is possible to introduce an effective 2D self-energy in the momentum representation and sum up the corresponding Dyson series.
The self-energy then reads:
\begin{gather}
  \Sigma(\bp_n) = -\delta\mu + \delta v (\bp_n\bm(\sigma))-\frac{i}{2\tau_0}-\frac{i\bn\bm{\sigma}}{2\tau_1},
\end{gather}
where $\bp_n$ is the effective 2D momentum defined right below Eq.~\eqref{green-zero}. Do not confuse it with thereal $p_y$, over which we integrated, when we introduced the potential~\eqref{potential2D}. We also define the effective two-dimensional scattering times $\tau_l$ ($l$ = 0, 1, 2) according to:
\begin{gather}
\label{times0}
    \begin{split}
        \frac{1}{2\tau_{l}}=n_{\rm imp }u_0^2\frac{\ve}{2v^2p_0^5}\int g_{2,p_{\ve}(\bn-\bn^\prime)}\cos^{l}(\theta-\theta^\prime)\frac{d\theta^\prime}{2\pi}.
    \end{split}
\end{gather}
Here, we have $p_\ve=\ve/v$ and $\bn^{(\prime)}=(\cos\theta^{(\prime)}, \sin\theta^{(\prime)})$.

The very fact that the whole physics of the problem can be reformulated in terms of the 2D potential has a beautiful physical interpretation.
Let us recall that in the Landau gauge~\eqref{vector-potential}, the center of the orbit is given by $p_yl_H^2$. That is, the effective scattering rates~\eqref{times0} entering the perturbation theory are essentially ordinary scattering rates but averaged  over the positions of the center of the Landau orbit. With these perturbative building blocks, we are ready to compute the conductivity tensor.

\subsection{General expressions for  conductivities}
The conductivity, in the leading order of the expansion parameter ~\eqref{expansion}, is given by the following simple expression:
\begin{gather}
\label{cond-one}
    \sigma_{x,x[y]}=\frac{e^2\Omega^2}{4\pi^2v}\sum\limits_n\int {\rm Re}\big[{\rm Im}\big]\langle G^R_{n,11}G^A_{n+1,22}\rangle\frac{df}{d\ve}\,d\ve dp_z .
\end{gather}
Here, we discard the $G^RG^R$ and $G^AG^A$ terms as subleading in the $1/(\tau\ve)$ disorder expansion. Also, by $\sigma_{xy}$ we mean the normal part $\sigma^{\rm I}_{xy}$ of the Hall conductivity (see subsection IV.C below for the full computation of the Hall conductivity).

We switch from the summation over LLs to integration over $n$ (in
the limit $T,\mu\gg \Omega$, all the functions are smooth functions of $n$), we substitute $1=dn=v^2p_ydp_y/\Omega^2$, and turn to polar coordinates: $p_ydp_ydp_z=p^2\sin\theta d\theta dp$. Now, we need to find the nonperturbative vertex renormalization responsible for the difference between $\langle G^RG^A\rangle$ and $\langle G^R\rangle\langle G^A\rangle$.

As shown in the Appendix, we are able to perform the integration over the modulus of the momentum $p$ in~\eqref{cond-one} and end up with only an angular integral. As a result, the conductivity tensor~\eqref{cond-one} can be rewritten in the following form:
\begin{gather}
    \label{sigma-angle}
    \sigma_{x,x[y]} = \int\frac{d\ve}{2\pi}\frac{df(\ve)}{d\ve}\int\frac{d\theta\sin\theta}
    {2\pi}\Tr\{\Gamma^{RA}_x(\theta){\tilde \sigma}_{x[y]}\},
\end{gather}
where $\Gamma_x(\theta)$ is the so-called angular vertex function
\begin{gather}
\label{vertex-2D}
    \begin{split}
    &\Gamma_x^{RA}(\theta)=\sum\limits_n\int\frac{dp_zdp_y}{(2\pi)^2}\\
    &\times\delta\Big(\cos\theta-\frac{p_z}{p}\Big)\langle G^R_{x,x^\prime}(\ve,\bp)\,{\tilde \sigma}\,G^A_{x^\prime,x}(\ve,\bp)\rangle\,dx.
    \end{split}
\end{gather}
It is essentially a vertex function, integrated over the modulus of the momentum at fixed $v p_x/\ve$ ratio. The notation ${\tilde \sigma}_{x,y}$ in the r.h.s. of Eq.~\eqref{sigma-angle} stand for Pauli $\sigma$ matrices.
Then, we plugin the vertex expressions from~\eqref{vertex-first} and take the angular integral to obtain:
\begin{gather}
   \label{cond2D}
    \begin{pmatrix}
        \sigma_{xx}\\
        \\
        \sigma^{\rm I}_{xy}
    \end{pmatrix}=\frac{e^2}{2\pi v}\int\frac{d\ve\ve^2}{2\pi}\frac{df}{d\ve}\frac{\ve^2\tau_{\rm tr}(\ve)}{\tau_{\rm tr}(\ve)^2\Omega^4+\ve^2}
    \begin{pmatrix}
        1\\
        \\ \Omega^2\tau_{\rm tr}/\ve    \end{pmatrix}.
\end{gather}
Here, $\tau^{-1}_{\rm tr}\equiv \tau_0^{-1}-\tau_2^{-1}$ is the transport scattering rate.
Equation~\eqref{cond2D} is quite an important result. It shows that {\it in the long-range disorder limit, the conductivity is effectively recast in terms of the 2D Drude-type expression}.

Similar formulae for the $\delta$-correlated disorder were obtained in Ref.~\onlinecite{KlierPRB2017}. However, the magnetoconductance in Ref.~\onlinecite{KlierPRB2017} is expressed in terms of the 3D scattering rates. This is somewhat predictable, since the $\delta$-corrrelated disorder has zero correlation length and the scattering rate is not affected by any other scale, including the magnetic length, which is responsible for the change in the geometry of the problem.

For the disorder of the general type, with the correlation radius independent of the characteristic energy of the host system, we have for the transport scattering time:
\begin{gather}
   \label{time_tr}
   \tau^{-1}_{\rm tr}(\ve)=\frac{u_0^2T_{\rm imp}^3}{g_1\ve^2(p_0v)^2},\ \ \ T_{\rm imp}=n_{\rm imp}^{1/3}v.
\end{gather}
Here, $g_1=\int _0^\infty g(x^2)x^2\,dx$ is the numerical constant determined by the type of disorder. This way, we obtain  the {\it general expression for the longitudinal conductivity} for any relation between the chemical potential and temperature:
\begin{gather}
\label{cond-gen}
    \begin{split}
    \sigma_{xx}&=\frac{e^2}{g_1v}\frac{u_0^2T^2 T_{\rm imp}^3}{(p_0v)^2\Omega^4}f\left(\frac{T}{\tau_{\rm tr}(T)\Omega^2}\right),\ \ \hbox{where}\\
    f(a)&=\frac{4\pi^2}{3}-4\Big(a^2-\frac{\mu^2}{T^2}\Big)+\frac{2a^3}{\pi}{\rm Re}\left[\psi^\prime\left(\frac{1}{2}+\frac{a+i\mu/T}{2\pi}\right)\right],\\
    a &= \ \ \ \ \frac{u_0^2 T_{\rm imp}^3}{g_1(p_0v)^2\Omega^2 T}.
    \end{split}
\end{gather}
Here, $\psi(x)$  is the Euler's digamma function. The dimensionless parameter $a$ plays the role of the relative strength of the disorder.

The exact formula~\eqref{cond-gen} can be somewhat simplified by the interpolation expression (which becomes exact in the limit $\Omega\rightarrow0$ and $\Omega\rightarrow\infty$) making it more useful for experimental purposes:
\begin{gather}
\label{cond-appr}
\sigma_{xx}=\frac{e^2}{v}\tau_{\rm tr} \max\left\{T^2,\mu^2\right\}\left[1+\frac{7\pi^2}{5}\frac{\Omega^4\tau_{\rm tr}^2}{\max\left\{T^2,\mu^2\right\}}\right]^{-1}.
\end{gather}
Here, the transport scattering time $\tau_{\rm tr}$ should be taken at the energy $\ve=\max\{\mu, T\}$.
\begin{figure}[t]
\centering
 \includegraphics[width=0.4\textwidth]{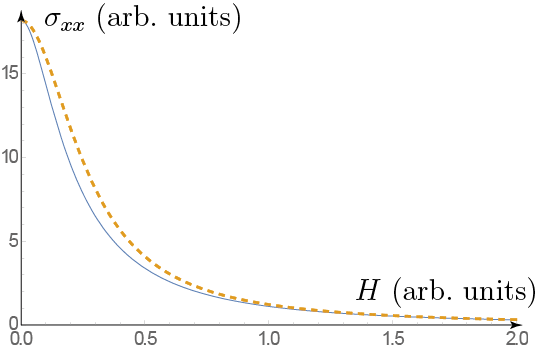}
 \caption{ (Color online) Conductivity $\sigma_{xx}$ given by the exact Eq.~\eqref{cond-gen} (solid line) and by the approximate Eq.~\eqref{cond-appr} (dashed line) expressions as a function of  dimensionless parameter $\Omega^2\tau_{\rm tr}/T\sim H.$}
\label{magneto}
\end{figure}
To give the reader an idea of how well the interpolation formula represents the exact result~\eqref{cond-gen}, we plot it in Fig.~\ref{magneto}.
Equations (\ref{cond-gen}, \ref{cond-appr}) reproduce the $T^4$ dependence at $\Omega\rightarrow0$, obtained in Ref.~\onlinecite{Sarma2015} in the zero-field limit. The interpolation formula ~\eqref{cond-appr} for the conductivity $\sigma_{xx}$ effectively recasts it in the form of familiar Drude-type metallic expression
\begin{gather}
\sigma\propto \tau_{\rm tr}(1+\omega_c^2\tau^2_{\rm tr})^{-1},
\end{gather}
where $\omega_c=\Omega^2/2\ve$ is the semiclassical cyclotron frequency at energy $\ve$.

The same kind of Drude representation but with 3D scattering times was obtained in Ref.~\onlinecite{KlierPRB2017} for $\delta$-correlated disorder.

\begin{figure}[t]
\centering
 \includegraphics[width=0.45\textwidth]{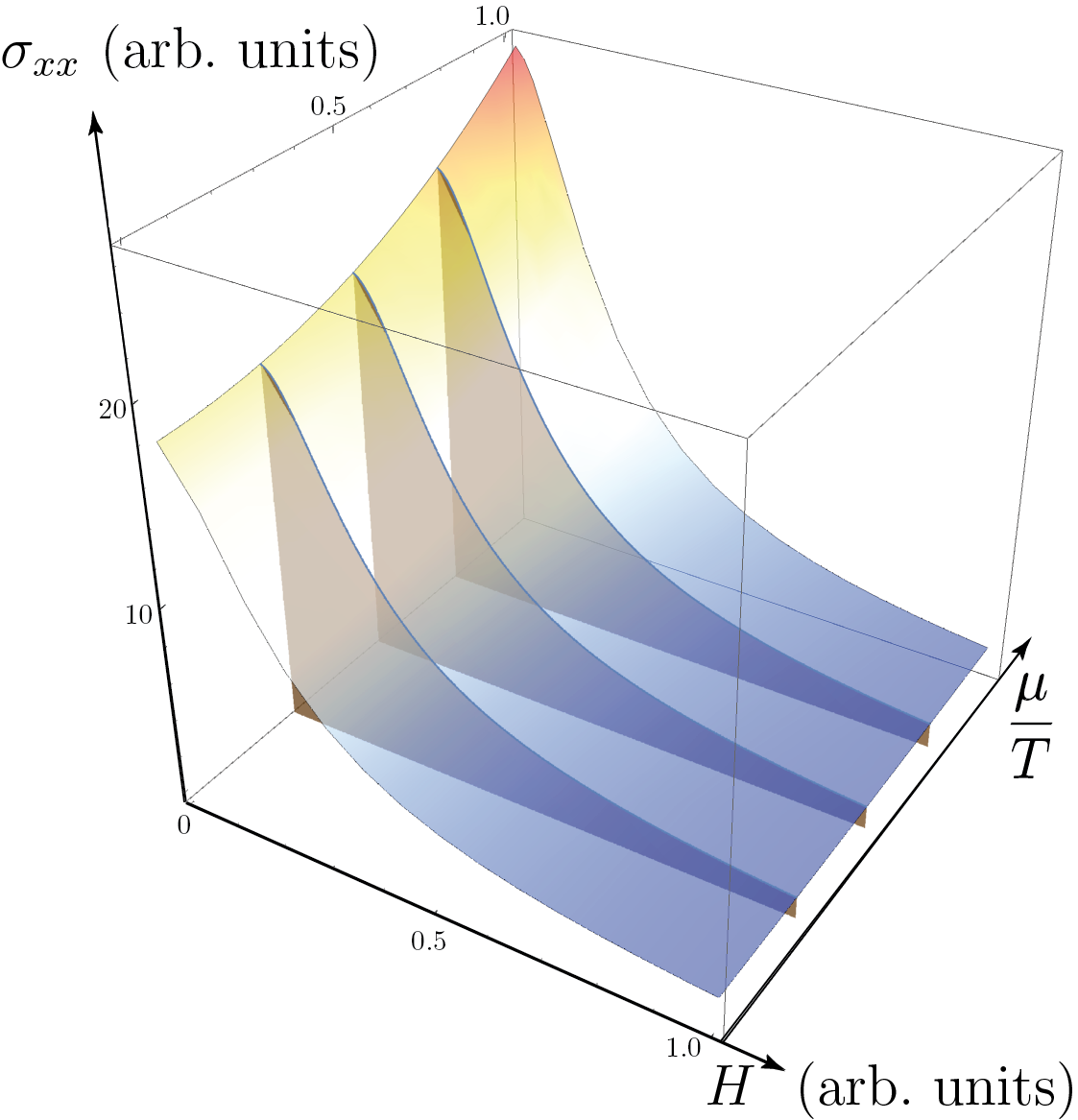}
 \caption{(Color online) Conductivity $\sigma_{xx}$ given by Eq.~\eqref{cond-gen} as a function of $\mu/T$ and the dimensionless parameter $\Omega^2\tau_{\rm tr}/T\sim H$.  Here we assume that $\mu\ll T$.}
\label{magneto1}
\end{figure}

The conductivity $\sigma_{xx}(H)$ for different values of $\mu/T$ are shown in Fig.~\ref{magneto1}.

\begin{figure}[t]
    \centering
    \includegraphics[width=0.4\textwidth]{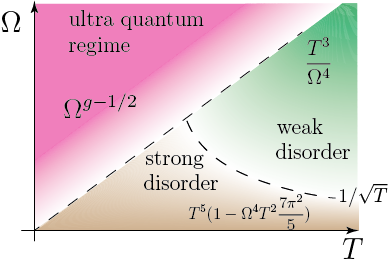}
    \caption{(Color online) Phase diagram for the conductivity $\sigma_{xx}$ for the non-Coulomb disorder, see Eqs.~\eqref{sigmaxx} and \eqref{cond-gen}. Here, $g=0$ for $\gamma\leq0$, and $g=\gamma/2$ for $0<\gamma<1$. See detailed explanations in the text.}
    \label{phase1}
\end{figure}

The behavior of the conductivity $\sigma_{xx}$ can be conveniently shown in the phase diagram (see Fig.~\ref{phase1}). The upper left red corner of this phase diagram corresponds to the ultra-quantum regime, $T\ll\Omega$, where depending on the characteristic exponent $\gamma$ of the impurity potential, we expect a $\gamma$-dependent scaling of $\sigma_{xx}$. The lower right corner is divided into the regimes of weak and strong disorder. The brown area corresponds to a strong disorder, and is described by Eq.\eqref{cond-appr} in the
\begin{gather}
\tau_{\rm tr}\ll \max\{T,\mu\}/\Omega^2
\end{gather}
limit. One could also refer it to as a weak magnetic field regime, where $\sigma_{xx}$ exhibits predominantly the $T^4$ dependence characteristic of a zero-field system with a correction proportional to $H^2$. The green area depicts the opposite weak-disorder limit (or that of high magnetic field), where the transport of charge carriers is strongly affected by the magnetic field.

Next, we calculate the Hall conductivity.

\subsection{Hall conductivity $\sigma_{xy}$}

As usual, the Hall conductivity is split into two parts: anomalous and normal ones. Let us first calculate the anomalous part. As before, we make use of~Eq.~\eqref{sigma-anom}. To regularize the expression for the charge-carrier density, we subtract the respective density at zero chemical potential, thus eliminating the contribution of the Fermi sea.
\begin{gather}
    \begin{split}
    n_0(H,\mu,T)
    =\frac{\Omega^2}{4\pi^2v^2}\int\limits_{-\infty}^\infty\frac{dp_z}{2\pi}  \sum\limits_n \Big(f(\ve_n-\mu)-f(\ve_n+\mu)\Big).
    \end{split}
\end{gather}

In the $\Omega\ll (T,\mu)$ limit, we use the Euler--MacLaurin summation formula:
\begin{gather}
    \sum\limits_{n=0}^\infty F(a+n)\approx\frac{F(a)}{2}+\int\limits_a^\infty F(x)\,dx.
\end{gather}
Then, we obtain:
\begin{gather}
    \label{density2}
    n_0(\mu,T) =\frac{1}{4\pi^2v^3}\left(\Omega^2\mu+\frac{2}{3}\big[\mu^3
    +\pi^2\mu^2T\big]\right).
\end{gather}
Only the first term in Eq.~\eqref{density2} is field dependent.
Despite its smallness, it is this term that contributes to $\sigma^{\rm II}_{xy}$.
This way, we arrive at the following expression:
\begin{gather}
   \label{sigmaxy2}
    \sigma_{xy}^{\rm II}=\frac{e^2\mu}{2v}.
\end{gather}
In experiments, the charge-carrier density is constant for each sample of WSM. Hence, the chemical potential is almost field independent in the high-temperature regime (see Discussion section for the relevant estimates).

Next, we calculate the normal part of $\sigma^{\rm I}_{xy}$ given by~\eqref{cond2D}. The conductivity $\sigma^{\rm I}_{xy}$ can be computed exactly at any value of $\mu$ (see the corresponding integral derived in the Appendix):
\begin{gather}
\label{sigmaxy-ex}
    \begin{split}
    &\sigma_{xy}^{\rm I} = \frac{e^2T^2\mu}{\pi^2 v\Omega^2}f_1(a),\\
     &f_1(a)= \frac{\mu^2}{T^2}+\pi ^2-a^2
   - \frac{a^4 T}{2\pi\mu}{\rm Im}\left[\psi ^{(1)}\left(\frac{a}{2 \pi }+\frac{1}{2}+\frac{i \mu/T }{2 \pi }\right)\right],
    \end{split}
\end{gather}
where $a$ is defined in Eq.~\eqref{cond-gen}.

As before, we concoct a Drude-type interpolation formula from $\Omega\rightarrow 0$ to $\Omega^2\gg\tau_{\rm tr}^{-1}\max\{T,\mu\}$ using formula~\eqref{sigmaxy-ex}:
\begin{gather}
  \label{sigmaxy-appr}
   \sigma_{xy}^{\rm I} \approx \frac{c_1e^2\mu}{v}\frac{ \Omega^2\tau_{\rm tr}^2}{1+\frac{c_2\tau_{\rm tr}^2\Omega^4}{\max\{T^2,\mu^2\}}},
\end{gather}
where $c_{1,2}=1$ if $\mu\gg T$ and $c_{1,2}=7\pi^{2(4)}/3$ if $\mu\ll T$. Here $\tau_{\rm tr}(\ve)$ is taken at $\ve=\max\{\mu, T\}$.
Depending on the factor $\omega\tau_{\rm tr}$, the normal part may be a leading or subleading contribution to the Hall conductivity. Qualitatively, $\sigma_{xy}$ can be described by the following identity:
\begin{gather}
   \sigma_{xy}=\frac{e^2\mu}{v}\begin{cases}
   d_1,\ \ & \Omega\tau_{\rm tr}\ll 1,\\
   \\
   d_2\Omega^2\tau^2_{\rm tr},\ &\ 1\ll\Omega\tau_{\rm tr}\ll \frac{\max\{T,\mu\}}{\Omega},\\
   \\
    d_3\frac{\max\{T^2,\mu^2\}}{\Omega^2},\ &\  \frac{\max\{T,\mu\}}{\Omega}\ll \Omega\tau_{\rm tr}.
                                \end{cases}
\end{gather}
Here $d_i$ are numerical factors following from relations~\eqref{sigmaxy2} and \eqref{sigmaxy-appr}.

\begin{figure}[h]
\centering
 \includegraphics[width=0.4\textwidth]{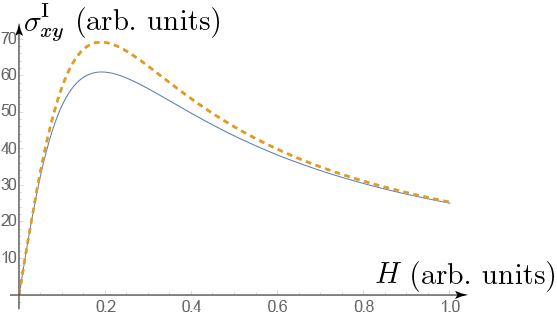}
 \caption{(Color online) Hall conductivity $\sigma_{xy}$ given by Eq.~\eqref{sigmaxy-ex} (solid line) and by Eq.~\eqref{sigmaxy-appr} (dashed line) as a function of the dimensionless parameter $\Omega^2\tau_{\rm tr}/T\sim H$.}
\label{magnetoxy}
\end{figure}

The plot illustrating the accuracy of the interpolation formula~\eqref{sigmaxy-appr} is presented in Fig.~\ref{magnetoxy}.

Interestingly,  the case of Coulomb impurities does not lead to different results, despite the fact that the Debye radius depends on temperature. The relatively slow decay of the Coulomb correlation function leads to a trivial logarithmic enhancement of the respective 2D scattering rate:
\begin{gather}
    \label{time_c}
  \tau^{-1}_{\rm tr} =2\pi^2\frac{\alpha^2 T^3_{\rm imp}}{\ve^2}\ln\frac{1}{\alpha}.
\end{gather}
Otherwise, the whole temperature and field dependence remains the same.

\begin{figure}[h]
\centering
\includegraphics[width=0.45\textwidth]{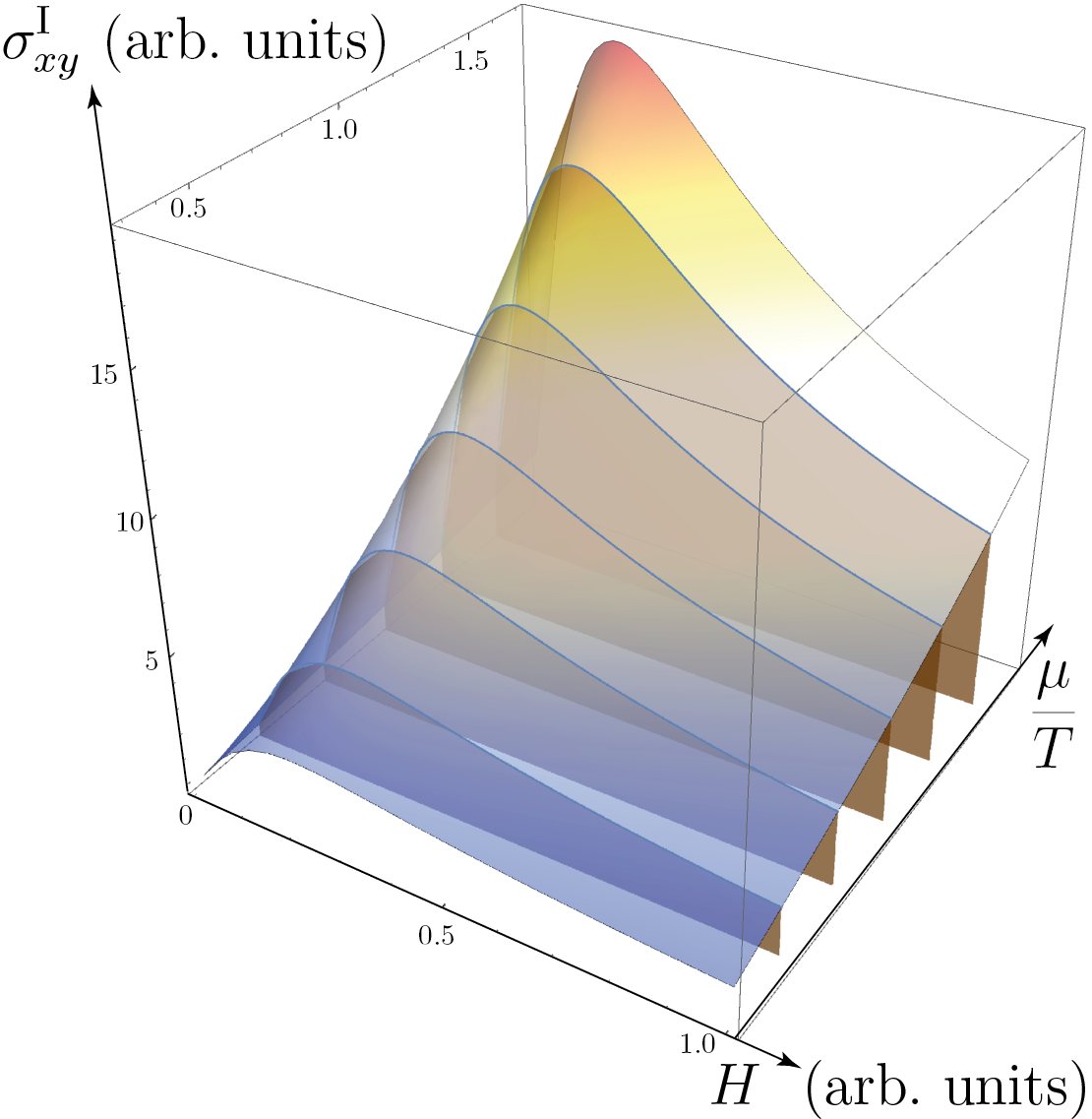}
 \caption{(Color online) Conductivity $\sigma^{\rm I}_{xy}$ given by Eq.~\eqref{cond-gen} as a function of $\mu/T$ and the dimensionless parameter $\Omega^2\tau_{\rm tr}/T\sim H$.}
\label{sigmaxy1}
\end{figure}

The plot of the Hall conductivity $\sigma_{xy}$ is presented in Fig.~\ref{sigmaxy1} as a function of the magnetic field $H$ and chemical potential $\mu$.

\section{Discussion}
\label{Discussion}

In this paper we have performed a detailed analysis of the effect of the long range disorder introduced by impurities on the magnetotransport in WSMs. Our study is mainly focused on the magnetic field and temperature dependence of the transverse magnetoconductivity. Two important limiting cases are considered: ($i$) the ultra-quantum limit, corresponding to low temperatures (or high magnetic field), for which the main contribution comes from the zeroth Landau level, and ($ii$) the opposite semiclassical limit, when a large number of Landau levels is involved in the transport phenomena.

We have completely discarded the effects  of the internode charge transfer. In principe, this effect can be important at sufficiently high fields
as argued in Ref.~\onlinecite{Tikh2019}. The necessary condition for the applicability of our approach is $\tau^{-1}_{\rm inter}\ll \tau^{-1}_{\rm intra}$, where for finding the scattering rates, we can use e.g.  Eqs.~\eqref{times0} or~\eqref{time_tr}. As derived in Ref.~\onlinecite{Tikh2019}, this condition is equivalent to $H\ll\alpha^{-3/2} eQ^2/v$, where $Q$ is the distance between the Weyl nodes in momentum space.

However, for a typical WSM like TaAs~\cite{XuScience2015,Lv2015,Ramshaw2018}, we extract the separation between Weyl nodes as $Q=0.01$ \AA$^{-1}$, while the Fermi velocity $v\approx 3\times 10^5$ m/s, which gives the respective field estimate of $H\sim 50$ T even for the fine structure constant in TaAs (unknown to us at the moment) $\alpha\sim 1$. Therefore, we safely discard this effect.

The long-range impurity potential is chosen in a rather general form. We show that in the ultra-quantum limit the nonlinear magnetoresistivity dependence
on the magnetic field is the manifestation of the singular (non-Coulomb) short-range behavior of the impurity potential and its correlation function.
In the semiclassical limit, we have demonstrated that unlike the short-range disorder case~\cite{KlierPRB2017}, the long-range disorder makes the scattering in the system essentially two-dimensional. We derived general formulae for $\sigma_{xx}(H)$ and $\sigma_{xy}(H)$ valid within a wide range of values of temperature and chemical potential.

In typical experiments~\cite{NovakPRB2015}, the doping levels in WSMs are rather high ($\sim10$ meV), which corresponds to $\mu~\sim 100$ K. The typical magnetic fields $H\sim 1$ T correspond to the gap between the zeroth and first Landau levels $\sim 10$ K. Reference~\onlinecite{Ramirez}, however, reports the observation of WSM in an almost undoped regime $\mu\ll T$. Therefore, both  the $\mu\ll T$ and $\mu\gg T$ regimes seem experimentally viable and the relation between $\mu,\ H$, and $T$ can be quite general. Hence, our results obtained in both limits can be relevant.

We can also mention the numerical work in Ref.~\onlinecite{PLeePRB2015}. In this paper, Coulomb impurities are correctly identified as the long-range disorder, and the high-temperature limit is explored. The nontrivial result of Ref.~\onlinecite{PLeePRB2015} is the scaling of the magnetoconductance $\sigma_{xx}\propto H^{-5/3}$ in the low-field regime $\Omega\lesssim T$.
Our analytical study addresses the case $T\gg\Omega$, and the aforementioned regime cannot be accessed in our semiclassical computation, where $\Omega/T\ll 1$ is the essential expansion parameter.

In a semiclassical regime $T\gg \Omega$, our results match the results reported in Ref.~\onlinecite{KlierPRB2015} in the low impurity concentration regime, $\tau_{\rm tr}\gg T\Omega^{-2}$, but differ in the opposite limit. We attribute this to the effect of the long-range disorder correlations.

Note also in conclusion that the problem under study has a deep analogy with the detailed analysis of the effect of interplay of quantum interference and disorder on magnetoresistance in the systems with the hopping conductivity undertaken in Refs~\onlinecite{LinNoriPRL1996,LinNoriPRB1996a,LinNoriPRB1996b}.

\begin{figure}[ht]
\centering
 \includegraphics[width=0.25\textwidth]{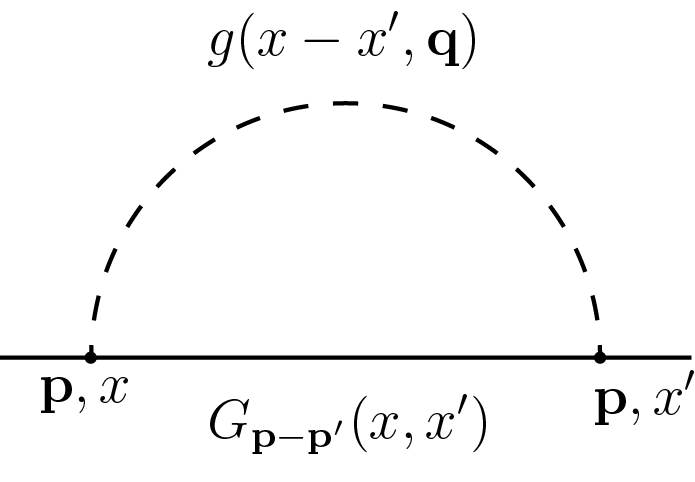}
 \caption{Electron self-energy}
\label{fig4}
\end{figure}

\acknowledgments

This work is partly supported by the Russian Foundation for Basic Research, project Nos. 19-02-00421, 19-02-00509, 20-02-00015, and 20-52-12013, by the joint research program of the Japan Society for the Promotion of Science and the Russian Foundation for Basic Research, JSPS-RFBR Grants No. 19-52-50015 and No. JPJSBP120194828, and by Deutsche Forschungsgemeinschaft (DFG), grant No. EV 30/14-1. F.N. is supported in part by NTT Research, Army Research Office (ARO) (Grant No. W911NF-18-1-0358), Japan Science and Technology Agency (JST) (via the CREST Grant No. JPMJCR1676), Japan Society for the Promotion of Science (JSPS) (via the KAKENHI Grant No. JP20H00134), and Grant No. FQXi-IAF19-06 from the Foundational Questions Institute Fund (FQXi), a donor advised fund of the Silicon Valley Community Foundation. Ya.I.R. acknowledges the support of the Russian Science Foundation, project No. 19-42-04137.
\appendix

\begin{widetext}
\section{Perturbation theory for $T\gg\Omega$}
\subsection{Self-energy and Dyson series for the Green's function}

 The diagram representing the first-order correction to the Green's function is presented in Fig.~\ref{fig4} and is given by the following analytical expression
\begin{gather}
\label{self-energy}
    \begin{split}
       &\delta G(x,x^\prime)=\sum_{n,m,k}\int\frac{dq_xdq_ydq_z}
       {(2\pi)^3}e^{iq_x(x^1-x^2)}g(q_x,q_{yz})\\
       &\times S_{n}(x_{p_y})G_n(p_z)S^\dag_{n}(x^1_{p_y})S_{m}(x^1_{p_y+q_y})
       G_m(p_z+q_z)S^\dag_{m}(x^2_{p_y+q_y})
    S_{k}(x^2_{p_y})G_k(p_z)S^\dag_{k}(x^\prime_{p_y}).
    \end{split}
\end{gather}
The correction to the Green's function is a matrix, each term containing a product of $\chi_{l}(x_{p_y})$ functions.

It is essential that for large $n$ we discard the difference between $n$ and $n+1$, when computing the correlation function. This allows us to treat the matrices $S_{n}(x_{p_y}),\ S^\dag_{n}(x_{p_y})$ as proportional to unit ones: $S_{n}(x_{p_y}),\ S^\dag_{n}(x_{p_y}) = \chi_{n}(x_{p_y})\cdot\bm{1}$.
The leading contribution to the conductivity comes from $n\gg1$ (in fact, we will later see that the contribution comes from $n\sim (T/\Omega)^2$) and use the asymptotic relation for the Hermite polynomials~\cite{AbrStegun}:
\begin{gather}
\label{Hermite}
   e^{-x^2/2}H_n(x)\approx\Big(\frac{2n}{e}\Big)^\frac{n}{2}\sqrt{2}
   \cos\left(x\sqrt{2n}-\frac{n\pi}{2}\right)\left(1-\frac{x^2}{2n+1}
   \right)^{-\frac{1}{4}}.
\end{gather}
The question is how to simplify the product
\begin{gather}
\chi_m(x^1_{p_y+q_y}) \chi_m(x^2_{p_y+q_y})e^{iq_y(x^1-x^2)}\,dq_y
\end{gather}
entering~\eqref{self-energy}. We split the product of two cosine functions in each Hermite polynomial into
\begin{gather}
 \cos[\sqrt{2m}(x_{p_y}^1-x_{p_y}^2)/l_H]+\cos(\sqrt{2m}[(x_{p_y}^1+x_{p_y}^2)/l_H-2q_y l_H].
 \end{gather}
Then, we perform integration over $q_y$. The first term gives just the integral
\begin{gather}
g(q_x,q_{yz})dq_y/2\pi=g_{2,xz}.
\end{gather}
It is simply an effective 2D potential~\eqref{potential2D}. The second term is proportional to $U(\sqrt{2m}l_H, q_{yz})$. However, the correlation radius of the potential obeys the inequality $r_0\ll l_H\ll\sqrt{2m}l_H$. As a result, the term proportional to $\cos(\sqrt{2m}[(x_{p_y}^1+x_{p_y}^2)/l_H-2q_y l_H])$ is suppressed.

Now we are able to perform the next estimate:
\begin{gather}
    \label{inter1}
    \begin{split}
       &\delta G(x,x^\prime)\approx \sum_{n,m,k}\chi_{n}(x_{p_y}) \chi_{k}(x^\prime_{p_y})\int \frac{dq_z}{2\pi} G_n(p_z) G_m(p_z+q_z) G_k(p_z)\\
       &\times \int\frac{dq_x}{2\pi} e^{iq_x(x^1-x^2)}g_2(q_{xz})\frac{1}{\pi l_H\sqrt{2m}}\cos[\sqrt{2m}(x_{p_y}^1-x_{p_y}^2)/l_H]\frac{1}{\pi l_H}\left(\frac{4}{nk}\right)^{1/4}
    \frac{\cos[\sqrt{2n}x^1_{p_y}/l_H]}{\sqrt[4]{1-\frac{x^{1,2}_{p_y}}
    {2n}}}\frac{\cos[\sqrt{2k}x^2_{p_y}/l_H]}{\sqrt[4]{1-\frac{x^{2, 2}_{p_y}}{2k}}}dx^1 dx^2.
    \end{split}
\end{gather}
In Eq.~\eqref{inter1}, it is important to discern the difference between the fast-oscillating cosine-type terms in the nominator and slow algebraic factors in the denominator. To perform the integration over $x^1,\ x^2$,  we change $x^1,\ x^2\rightarrow r=x^1-x^2,\ x^2$.
We obtain many fast-oscillating terms (the relevant $n,\ m$ and $k$ are large). For example,  performing integration over $x^2$, we obtain:
\begin{gather}
    \label{inter2}
\begin{split}
      \frac{1}{2}\int dx^2  \frac{\cos[\sqrt{2n}r+(\sqrt{2n}+\sqrt{2k})x^2/l_H]+
      \cos[\sqrt{2n}r+(\sqrt{2n}-\sqrt{2k})x^2/l_H]}
      {\sqrt[4]{1-\frac{(r+x^{2}_{p_y})^2}{2nl_H^2}}
      \sqrt[4]{1-\frac{x^{2,2}_{p_y}}{2kl_H^2}}}.
\end{split}
\end{gather}
As we see from the structure of the integral of~\eqref{inter2}, the
nominator is a fast-oscillating function of $x^2_{p_y}$. As a result, the integral is suppressed unless $n=k$. Thus, the integral is $\propto \delta_{nk}$ in the main order for $1/\sqrt{n-k}$.
Let us compute it for $n=k$. The nominator does not oscillate any more, and we should analyze the denominator.
The important $r\sim p^{-1}_0$, which comes from $g_2(q_{xz})$. On the other hand, the important $x^2_{p_y}\sim\sqrt{n} l_H$. We see that $r\ll x^2_{p_y}$ for the denominator. Therefore, we integrate over $x^2_{p_y}$ trivially. We are then left with the following expression:
\begin{gather}
    \begin{split}
       &\delta G(x,x^\prime)\approx \sum_{n,m}\chi_{n}(x_{p_y}))\chi_{n}(x^\prime_{p_y})\int \frac{dq_z}{2\pi}  G_n(p_z)G_m(p_z+q_z)G_n(p_z)\\
       & \times \int\frac{dq_xdr}{2\pi} e^{iq_xr}g_2(q_{xz})\frac{1}{\pi l_H\sqrt{2m}}\cos(\sqrt{2m}r/l_H)\cos(\sqrt{2n}r/l_H).
    \end{split}
\end{gather}
While integrating over $r$, we obtain the combination of 4 $\delta$-functions. The only relevant ones are $\delta(q_x+\sqrt{2m}l_H^{-1}-\sqrt{2n}l_H^{-1})$ and $\delta(q_x-\sqrt{2m}l_H^{-1}+\sqrt{2n}l_H^{-1})$.
Consequently, we have the following final formula for the correction to the Green's function:
\begin{gather}
    \begin{split}
       \delta G(x,x^\prime)\approx \sum_{n,m}\chi_{n}(x_{p_y})\chi_{n}(x^\prime_{p_y})\int \frac{dq_z}{2\pi} G_n(p_z)G_m(p_z+q_z)G_n(p_z)\frac{1}{2\pi l_H\sqrt{2m}}g_2\left[(\sqrt{2m}-\sqrt{2n})l_H^{-1},q_{z}\right].
    \end{split}
\end{gather}
Using the fact that $\sqrt{n}l_H^{-1}\sim T\gg p_0$, we understand that $m$ in the last sum is actually very close to $n$. Indeed, we see that:
\begin{gather}
  \sqrt{m}-\sqrt{n}\sim p_0 l_H\gg1,\ \ {\rm while}\ \ \frac{\sqrt{m}-\sqrt{n}}{\sqrt{m}+\sqrt{n}}\sim \frac{p_0v}{T}\ll1.
\end{gather}
From the last inequalities, we see that the terms of the sum over $m$ are smooth functions of $m$ and the sum can be turned into an integral. Introducing the effective momentum $p^\prime_y=\sqrt{2m}l_H^{-1},\ \ dm\equiv 1 = q_ydq_y l_H^2$, and $p_z^\prime=p_z+q_z$, we obtain:
\begin{gather}
    \label{green-fin}
    \delta G(x,x^\prime)\approx \sum_{n}\chi_{n}(x_{p_y}) \left[\int \frac{d\bp^\prime}{(2\pi)^2} G(\bp_n)G(\bp^\prime)G(\bp_n)g_2(\bp^\prime-\bp_n)\right] \chi_{n}(x^\prime_{p_y}).
\end{gather}
Here, $\bp_n$ is introduced in~\eqref{green-zero}.
The expression in the square brackets in~\eqref{green-fin} allows us to build the ordinary 2D Dyson series for the Green's function as well as vertex functions determining the conductivity tensor. Indeed, it coincides with the standard expression of perturbation theory without magnetic field with the effective 2D potential $g_2(\bp^\prime-\bp_n)$.
Therefore, we can write the momentum-dependent self-energy as:
\begin{gather}
\label{self1}
   \Sigma(\bp_n) = \int \frac{d\bp^\prime}{(2\pi)^2} G(\bp^\prime)g_2(\bp^\prime-\bp_n).
\end{gather}

The resummed Green's function then reads:
\begin{gather}
    \label{green-fin1}
    G(x,x^\prime) \approx \sum_{n}\chi_{n}(x_{p_y}) [G^{-1}(\bp_n)-\Sigma(\bp_n)]^{-1} \chi_{n}(x^\prime_{p_y}).
\end{gather}
The resulting expression yields an irrelevant part, which can be absorbed into the renormalized chemical potential and Fermi velocity, and the dissipative part.
The imaginary part reads:
\begin{gather}
\label{self-im}
\begin{split}
    \Sigma^R(\bp_n)&=v.p.\int\frac{d\bp^\prime}{(2\pi)^2}
    \frac{\ve+\bp^\prime\bm{\sigma}}{(\ve+i0)^2-\bp_n^2}
    g_2(\bp^\prime-\bp_n)-i\pi\int \frac{d\bp^\prime}{(2\pi)^2}\left(\delta(\ve-\ve_n)+
    \delta(\ve+\ve_n)\right)\left[\frac{1}{2}+
    \frac{\bp^\prime\bm{\sigma}}{2\ve}\right]g_2(\bp^\prime-\bp_n)\\
    &=-\delta\mu + \delta v \bp_n\bm{\sigma}-\frac{i}{2\tau}-\frac{i\bn\bm{\sigma}}{2\tau_1},\\
    \frac{1}{\tau}&=\frac{p_n}{4\pi}\int g_2(\bn\bn^\prime)d\bn^\prime,\quad \quad \frac{1}{\tau_1}=\frac{\ve}{4\pi}\int (\bn\bn^\prime)g_2(\bn\bn^\prime)d\bn^\prime.
\end{split}
\end{gather}

\subsection{Vertex renormalization and conductivity tensor}
The Dyson equation for the vertex is built in a more subtle way.
In this case, the built-in magnetic anisotropy of the problem takes its toll. What we are going to do now is to introduce a slightly unusual definition of the vertex. We define the mass-shell vertex according to the following equation:
\begin{gather}
    \Gamma_x(\theta)=\sum\limits_n\int\frac{dp_zdp_y}{(2\pi)^2}
    \delta\Big(\cos\theta-\frac{p_z}{p}\Big)\langle G^R_{x,x^\prime}(\ve,\bp)\,\sigma_x\,G^A_{x^\prime,x}(\ve,\bp)\rangle\,dx.
\end{gather}
Then, the conductivity tensor assumes the form in Eq.~\eqref{sigma-angle}. In the zeroth-order ladder approximation, we change $\langle G^R G^A\rangle = \langle G^R\rangle\langle G^A\rangle$, and the vertex becomes:
\begin{gather}
\begin{split}
\label{vertex-zero}
   &\bm{\Gamma}^0_{x}
    =
   \frac{1}{2\pi l_H^2}\sum\limits_{n}\frac{dp_z}{2\pi}
   \begin{pmatrix}
        0 \ &\ G^R_{11,n}G^A_{22,n+1}\\
        \ G^R_{22,n}G^A_{11,n-1}\ &\ 0
   \end{pmatrix}.
\end{split}
\end{gather}

Changing the sum and the integral using the semiclassical approximation
\begin{gather}
\sum_n\int\frac{dp_z}{2\pi} \delta\big(\cos\theta-\frac{p_z}{p}\big) =l_H^2\int\frac{p^2dp}{2\pi},
\end{gather}
we obtain:
\begin{gather}
  \Gamma^{RA,0}_{x}
    =
   \frac{n_y^2}{2\pi v^3}
   \begin{pmatrix}
        0 \ &\ [\frac{1}{\tau}+\frac{1}{\tau_1}+\frac{i\Omega^2}{\ve}]^{-1}\\
        \ [\frac{1}{\tau}+\frac{1}{\tau_1}-\frac{i\Omega^2}{\ve}]^{-1}\ &\ 0
   \end{pmatrix}.
\end{gather}
In the first order of perturbation theory, the picture changes slightly, and we obtain the first stair of the ladder series

\begin{gather}
\label{vertex-first0}
\Gamma^{RA,1}_{x}= \frac{n_y^2}{2\pi v^3}
    \begin{pmatrix}
        0 \ &\ \frac{\frac{1}{\tau_1}+\frac{1}{\tau_2}}
        {\left[\frac{1}{\tau}+\frac{1}{\tau_1}+
        \frac{i\Omega^2}{\ve}\right]^2}\\
        \ \frac{\frac{1}{\tau_1}+\frac{1}{\tau_2}}
        {\left[\frac{1}{\tau}+\frac{1}{\tau_1}-\frac{i\Omega^2}
        {\ve}\right]^2}\ &\ 0
   \end{pmatrix}.
\end{gather}
In the higher orders of the perturbation theory the pattern repeats itself. As a result, we are able to perform the full summation:
\begin{gather}
\label{vertex-first}
\Gamma^{RA}_{x}= \frac{n_y^2}{2\pi v^3}
    \begin{pmatrix}
        0 \ &\ \left[\frac{1}{\tau_{\rm tr}}+\frac{i\Omega^2}{\ve}\right]^{-1}\\
        \left[\frac{1}{\tau_{\rm tr}}-\frac{i\Omega^2}{\ve}\right]^{-1}\ &\ 0
   \end{pmatrix}.
\end{gather}
Finally, we are ready to obtain the expression for the conductivity:
\begin{gather}
    \sigma_{x,x[y]} = \int\frac{d\ve}{\pi}\frac{df(\ve)}{d\ve}\int\frac{d\theta\sin^3\theta}{2\pi}{\rm Re}[{\rm Im}]\left[\frac{1}{\tau_{\rm tr}}+\frac{i\Omega^2}{\ve}\right]^{-1}.
\end{gather}

%\pagebreak

\end{widetext}

\section{Calculation of the integral for the conductivity at $T\gg\Omega$}.

The integral that enters the upper matrix element in the expressions for the conductivities in ~\eqref{cond2D} has the following form:
\begin{gather*}
    I(a)=\int\limits_{-\infty}^\infty \frac{1}{x^2+a^2}\frac{x^4 dx}{\cosh^2\frac{x-\mu}{2}}\\
    =\int\limits_{-\infty}^\infty\frac{x^2+\mu^2-a^2}{\cosh^2\frac{x}{2}}dx+a^4
    \int\limits_{-\infty}^\infty\frac{1}{\cosh^2\frac{x}{2}}\frac{dx}{(x+\mu)^2+a^2}.
\end{gather*}
The last integral is equal to
\begin{gather}
        \int\limits_{-\infty}^\infty\frac{1}{\cosh^2\frac{x}{2}}\frac{dx}{(x+\mu)^2+a^2}=\frac{2}{\pi a}{\rm Re}\left[\psi^\prime\left(\frac{1}{2}+\frac{a-i\mu}{2\pi}\right)\right].
\end{gather}
As a result, we recover Eq.~\eqref{cond-gen}
In the same manner we perform the integration for $\sigma^{\rm I}_{xy}$ in the lower part of~\eqref{cond2D} to obtain the exact expression ~\eqref{sigmaxy-ex}.

\end{document}